\documentclass{WileyMSP-template}

\usepackage{xcolor}
\usepackage[colorlinks, linkcolor = blue,hypertexnames = false,citecolor = teal]{hyperref}

\usepackage{subcaption}
\usepackage{todonotes}

\begin{document}

\pagestyle{fancy}
\rhead{\vspace{1em}}
\lhead{}

\title{What do Large Language Models \textit{know} about materials?}

\maketitle

\author{Adrian Ehrenhofer*}
\author{Thomas Wallmersperger}
\author{Gianaurelio Cuniberti}

\begin{affiliations}
Dr.-Ing. Adrian Ehrenhofer\\
Institute of Solid Mechanics \& Dresden Center for Intelligent Materials, TU Dresden \\
George-Bähr-Str. 3c, 01069 Dresden \\
Email Address: adrian.ehrenhofer@tu-dresden.de \\

Prof. Dr.-Ing. Thomas Wallmersperger \\
Institute of Solid Mechanics \& Dresden Center for Intelligent Materials, TU Dresden \\
George-Bähr-Str. 3c, 01069 Dresden \\
Email Address: thomas.wallmersperger@tu-dresden.de \\

Prof. Dr. Gianaurelio Cuniberti\\
Institute of Materials Science \& Max Bergmann Center of Biomaterials, Technische Universität Dresden, 01062 Dresden, Germany \& Dresden Center for Intelligent Materials, TU Dresden \\
George-Bähr-Str. 3c, 01069 Dresden \\
Email Address: gianaurelio.cuniberti@tu-dresden.de
\end{affiliations}


\begin{abstract}
Large Language Models (LLMs) are increasingly applied in the fields of mechanical engineering and materials science. As models that establish connections through the interface of language, LLMs can be applied for step-wise reasoning through the Processing-Structure-Property-Performance chain of material science and engineering. 
Current LLMs are built for adequately representing a dataset, which is the most part of the accessible internet. However, the internet mostly contains non-scientific content. If LLMs should be applied for engineering purposes, it is valuable to investigate models for their intrinsic knowledge -- here: the capacity to generate correct information about materials. 
In the current work, for the example of the Periodic Table of Elements, we highlight the role of vocabulary and tokenization for the uniqueness of material fingerprints, and the LLMs' capabilities of generating factually correct output of different state-of-the-art open models. This leads to a material knowledge benchmark for an informed choice, for which steps in the PSPP chain LLMs are applicable, and where specialized models are required. 
\end{abstract}

\tableofcontents

\clearpage

\section{Introduction: Large Language Models as data sources}
Today's Large language models (LLMs) are built to generate cohesive and context-relevant tokens, whereas the relevance is determined from a training dataset \cite{Vaswani2017}. Today's common LLMs include most of the open internet, including various valuable and reliable sources for materials data. Furthermore, many scientific works -- often obtained through means of legally gray areas \cite{Carlini2022} -- build a backbone of a highly trustworthy dataset \cite{Gao2020a}. One of the most fascinating aspects in scientific/engineering applications of this kind of models is to identify, how they can be adequately included in the process of designing engineering solutions. If they are adequately accurate, mechanical engineers and materials scientists can include them in their design process \cite{Ehrenhofer2025smart_materials_informatics}. 

\begin{figure}[h]
	\centering
	\includegraphics[width=0.7\linewidth]{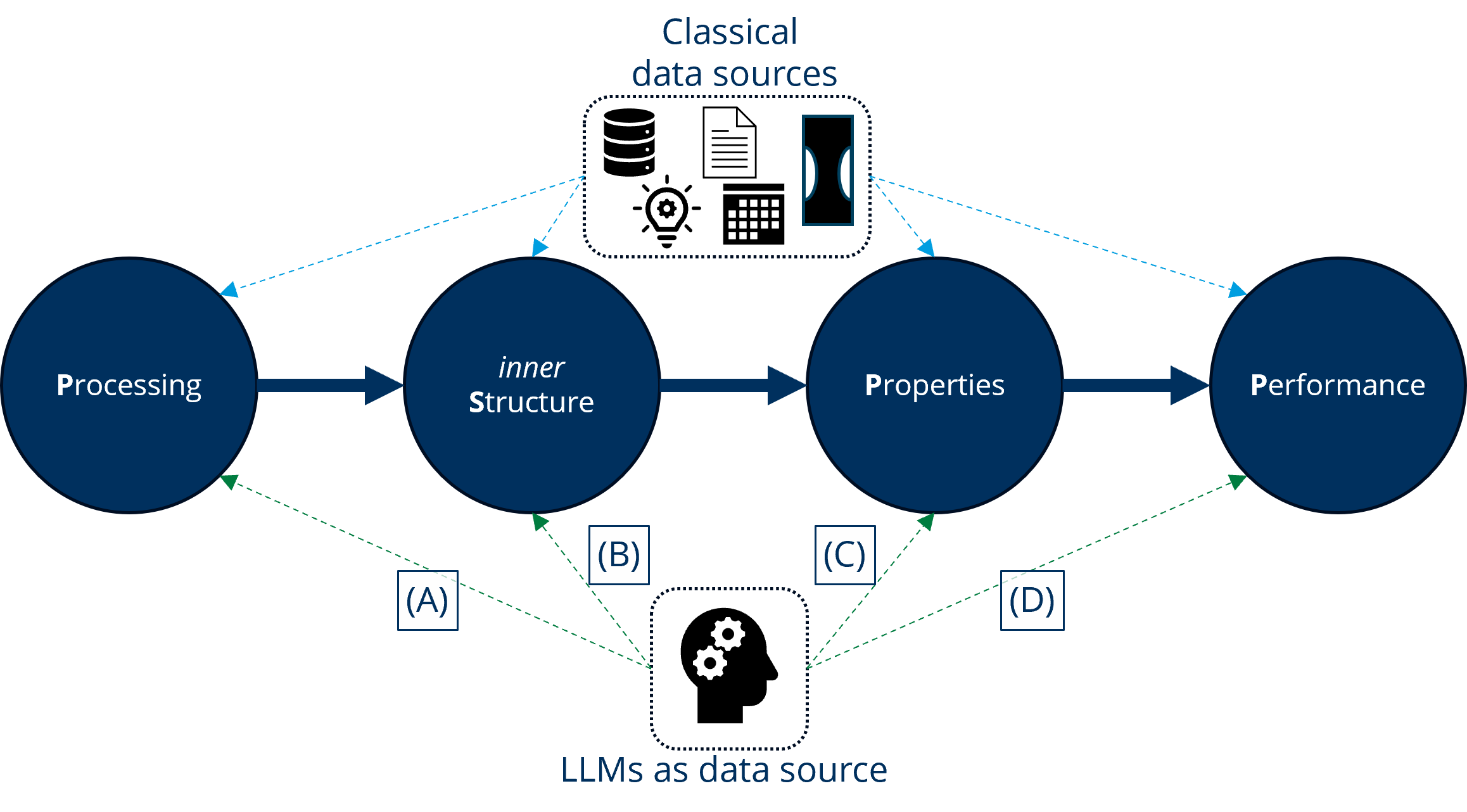}
	\caption{Classical datasources in the PSPP chain and possible applications for LLMs as data sources. If the complete chain should be filled in course of a PSPP chain reasoning process, the intrinsic knowledge for each step must be accurate. }
	\label{fig:intro_linear_end_to_end}
\end{figure}

In the current work, we will present how the inclusion of LLMs in an end-to-end engineering process can be realized. This can be understood as a reasoning process through the steps of the Processing-Structure-Property-Performance chain, whereas every step -- or the combination of multiple steps -- can be represented by models, see Figure \ref{fig:intro_linear_end_to_end}. In traditional implementations, these connections are typically made by physically-based models from a wide range of scientific and engineering disciplines, such as atomic simulation approaches with molecular dynamics, or continuum models for larger structures. According to the fourth paradigm of material science \cite{Agrawal2016,Wang2024,Curtarolo2013,Choudhary2022}, physically-based models can be replaced with machine-learning based models, such as LLMs. However, this is only possible, if these models are adequately \textit{knowledgeable} about materials. Based on the different inclusion points of LLM "knowledge" in the Processing-Structure-Property-Performance chain, different relevant output qualities must be evaluated to obtain information about the trustworthiness.

Various approaches for including machine learning in material science and engineering have been proposed, often under the name of the so-called fourth paradigm of materials science \cite{Agrawal2016}. The term \textit{Materials Informatics} \cite{Ramakrishna2019,Zhou2019} is often used in this context to highlight the use of computer science approaches for materials discovery. Specialized datasets, from the field of physics/chemistry \cite{Ohana2024,Levine2025} and of materials science \cite{Ong2013,Huan2016} are used to train end-to-end models for replacing the models that connect the PSPP relationship. Various methods allow the discovery of a wide range of material classes ranging from high entropy ceramics \cite{Curtarolo2012,Sarker2018} and polymers \cite{DoanTran2020,Kuenneth2021} to functional proteins \cite{Jumper2021} and composites \cite{Yin2021}. 

Autonomous (self-driving) labs \cite{Szymanski2023} further allow the automated synthesis and characterization of predicted materials to close gaps in material knowledge. Statistical methods for the planning of experiment allow the discovery of gaps in materials datasets, which can be closed by (automated) experiments to allow interpolation inside the design space, and to avoid extrapolation. 
\clearpage
An important aspect in uniquely identifying materials is the concept of the material fingerprint \cite{Huan2015,Ramprasad2017}, which is a set of descriptors that can includes its chemical composition and various additional aspects, such as the processing history and its parameters. Material fingerprints can be understood in analogy to embeddings in Large Language Models: For each material, they define a point in an abstract space, whereas materials can be understood as more similar, if their fingerprint has a small Eucledian distance. Adding more information into the fingerprint can thus serve a similar cause as creating embeddings from multiple modalities in vision-language models. The concept of the material genome is wider in its scope and also includes processes of materials development and high-throughput experimentation \cite{DoanTran2020,Xu2019}. 

One approach for fulfilling the promise of end-to-end models for material discovery is built on language \cite{White2023,Jablonka2023,Zhang2024b,Jablonka2024}. The quest of automating scientific discovery also leads to the application of LLMs for automated literature reviews \cite{Scherbakov2025} and the use of Retrieval-Augmented-Generation approaches for data-mining from papers: To obtain a material fingerprint from one or multiple databases, methods such as text and data mining can be applied \cite{Kononova2019,Dagdelen2024,OliveiraJr2022}. An important step in identifying chemical datasets for obtaining processing data is Named Entity Recognition (NER) and the construction of knowledge-graphs \cite{Bai2025}. In the next steps, data enrichment and synthetic data \cite{Patki2016} is important here.

\paragraph{Work overview} In Figure \ref{fig:intro_linear_end_to_end}, the steps of the PSPP chain for material discovery are presented. The classical data sources at every point are scientific papers/text books, material databases, patents, manufacturer's lists and individual experimental measurements (which can also be parts of any other of the source types). The question that we want to address in the current work is the suitability of replacing classical data sources with general-purpose Large Language Models. Therefore, in section \ref{sec:mat_met_pspp}, the addition of LLM knowledge in every step will be discussed. Then, in section \ref{sec:mat_met_llm_statistics}, the relevant background of LLMs and the applied statistical methods will be introduced and some details about the Periodic Table of Elements (PTE), which serves as a ground truth dataset, will be given in section \ref{sec:mat_met_pse}. Based on these foundations, several factors of material knowledge in LLMs will be presented as results in section \ref{sec:results_and_discussion}. This includes the question of uniqueness of tokens for chemical elements in section \ref{sec:res_unique_elements} and the quality of property prediction in section \ref{sec:res_property_prediction}. The conclusion is drawn in section \ref{sec:conclusion}.

\clearpage
\section{Application of Large Language Models in the Processing-Structure-Property-Performance chain}\label{sec:mat_met}
The current section provides the background to the chain of PSPP relationships (section \ref{sec:mat_met_pspp}) and how material knowledge from LLMs can be added to every step. The Language Models of interest (Generative Pretrained Transformer models), the significance of embeddings and possibilities for statistical analysis are specified in section \ref{sec:mat_met_llm_statistics}. Finally, the Periodic Table of Elements as the ground truth for the analysis is discussed in section \ref{sec:mat_met_pse}. 

In the current work, we define \textit{material knowledge} based on the broader concept of knowledge in LLMs and knowledge graphs \cite{Wang2025,Bai2025,OliveiraJr2022}. Furthermore, the following assumptions are taken:
\begin{itemize}
	\item The knowledge is specifically \textbf{linked to a material name} (i.e., a unique identifier). In this context, a material is defined by its specific \textbf{material fingerprint}, which includes the chemical constituents and all processing steps. Please note that, however, multiple names can describe the same material; this is especially relevant if producer's brand names or different nomenclatures for chemicals -- such as the CAS number, the PubChem CID, or the chemical formula -- are used to describe the same material. 
	\item \textbf{Context} is crucial for generation, as will be substantiated in section \ref{sec:mat_met_llm_statistics}. In the following sections, examples -- written in \texttt{special font} -- will be provided. The prompt/query is marked with "Q:". Please note that the answer "A:" ends after the maximum number of generated tokens, not at the learned end of sequence marker. Furthermore, we only consider short context: Chain-of-thought prompting \cite{Wei2022,Brown2020} or reasoning techniques are effective, because they produce tokens as context, such that useful next tokens are more probable, which often leads to better outcomes. For the results in section \ref{sec:results_and_discussion}, a maximum of 20 tokens or the end of sequence token (e.g., \texttt{<eos>} or a self-defined token \texttt{<.}) is chosen as the maximum, whichever comes first, to limit generation. 
	\item Material knowledge is \textbf{timeless}, i.e., there is no frequent update of processing steps. For example, a procedure for creating a polymer is adequately described and does not change over time. New processing leads to new materials, which are relevant as additions. 
\end{itemize}

\begin{figure}[h]
	\centering
	\includegraphics[width=0.9\linewidth]{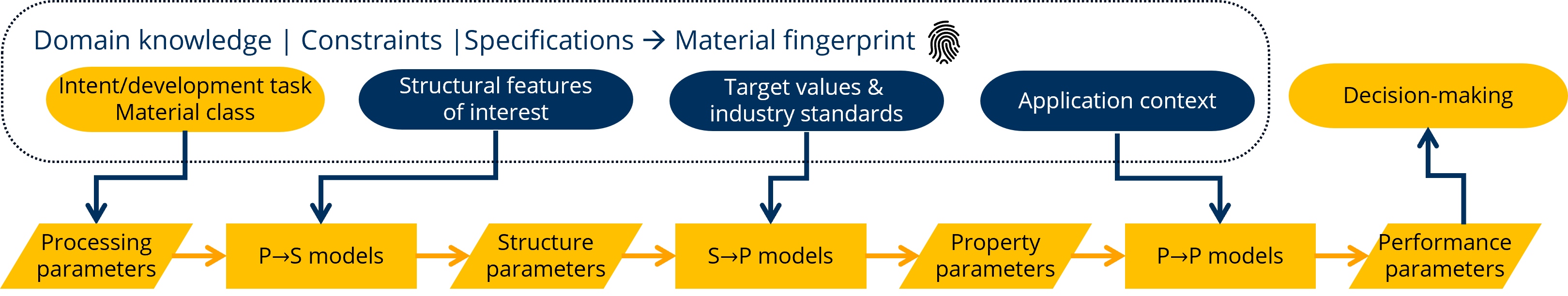}
	\caption{Reinterpretation of the chain from Figure \ref{fig:intro_linear_end_to_end} as a block flow diagram. Parallelograms represent datasets that are produced in the chain, while rectangular boxes are models that connect these nodes. The domain knowledge part is a set of inputs to the process, which are required. Together, they form something similar to a material fingerprint of a material inside an application context. In the data-driven approach of PSPP chain reasoning, each step is replaced by a LLM.}
	\label{fig:mat_met_process_chain}
\end{figure}

In Figure \ref{fig:mat_met_process_chain}, the above-provided PSPP-chain is written in the form of a block flow diagram, whereas the nodes are datasets that describe parts of the material fingerprint. Different types of models connect these data nodes; they depend on external constraints or specifications, which are part of the domain knowledge. In the data-driven approach, these models are replaced with blackbox machine learning models, e.g., LLMs. The appropriate context must be added to the prompt, as will be presented in the next subsection.

\renewcommand{\theenumi}{\Alph{enumi}}
\renewcommand{\labelenumi}{(\theenumi)}

\clearpage
\subsection{The PSPP-chain and ways to include LLM knowledge}\label{sec:mat_met_pspp}
At every single step of the PSPP chain (from (A) to (D) according to Figure \ref{fig:intro_linear_end_to_end}), information can be added. The steps will be explained in more detail. Since these steps are quite abstract, examples from the author's previous works in the field of macromolecules (smart hydrogels \cite{Ehrenhofer2025smart_materials_informatics} as parts of Soft-Hard Active-Passive Embedded Structures \cite{Ehrenhofer2025shapes_tables}) will be provided. The examples are specifically designed to show the kind of question and type of adequate answer that is possible from a LLM:
\begin{enumerate}
	\item From intrinsic material knowledge, LLMs can identify \textbf{processing steps} and valuable\footnote{Here, valuable means that they are significant for the material and that they have a numerical values. These aspects are discussed in previous works in the context of smart hydrogels \cite{Wang2024}.} processing parameters. Processing data output can only be generated if the processing steps are well-described in literature (in the data set) or if they can be logically derived as being important. In the same way as LLMs can generate cooking recipes, the creation of polymers can be generated \cite{White2023,Thik2023,Mavrogiannis2024}. However, hallucinations or misrepresentations (like adding glue to pizza \cite{Tan2024}) can happen, if this is represented in the data. An example for obtaining processing knowledge from LLMs is \newline 
	\texttt{Q:"For free radical polymerization of hydrogels, the following main ingredients are required: " \newline 
	A:"1. initiator 2. monomer 3. crosslinker 4. solvent 5..."\footnote{Generated with Llama-3.2-3B, Total output tokens: 20, only most probable token generated, i.e., temperature is zero (argmax selection). Generating more tokens will lead to additional ingredients. Please note that the blank space at the end of the prompt makes the beginning of a list more probable in the chosen model. As in the other examples, the generation stops mid-sentence because of the fixed number of generated tokens.}}
	\item LLMs can reproduce basic concepts of \textbf{inner structure} -- for example the granularity of metals or chain architecture of polymers -- and what they signify for resulting properties. This knowledge can be qualitative (impact of more coarse-grained to fine-grained or difference between statistical polymers and block copolymers) and may be represented in a quantitative fashion. Example \newline 
	\texttt{Q:"Compare two styrene-butadiene copolymers: the statistical variant mixes monomers randomly and shows one glass transition; the block variant" \newline 
	A:"mixes monomers in blocks and shows two glass transitions. The block variant is more brittle than the statistical..."\footnote{Generated with Llama-3.2-3B, Total output tokens: 20, argmax selection. The prompt was especially designed to provoke a fast reveal of the information within the first 20 tokens.}}. 
	\item \textbf{Property data} can be  gained from LLMs under the appropriate given context. As will be shown in section \ref{sec:mat_met_llm_statistics}, the context decides the logits and thus the probability distribution of tokens. The correct context can lead to the reproduction of correct output data. Results for this approach will be discussed in section \ref{sec:res_property_prediction}. In the example, a value for a very unspecific material is requested: \newline 
	\texttt{Q:"The elastic modulus of steel is " \newline 
	A:"200 GPa. The elastic modulus of concrete is 30 GPa. The elastic modulus of the..."\footnote{Generated with Llama-3.2-3B, Total output tokens: 20, argmax selection. Since no end of sequence token is generated, the model starts generating other property data.}}
	\item The \textbf{performance} of materials in a given environment -- e.g. as part of a smart composite in a Soft-Hard Active-Passive Embedded Structure \cite{Ehrenhofer2025} -- requires an adequate description of the application. LLMs must grasp the concept of the application due to intrinsic understanding of physical concepts or a continuum-based model with specific numerical output must be provided instead. In previous works \cite{Ehrenhofer2025smart_materials_informatics}, the example of a smart mesh in a helmet according to a previous invention was discussed \cite{Ehrenhofer2020spie_rainproofing,Binder2021pat_mesh}: The opening or closing performance of the pores can be judged by a YES/NO statement, depending on the maximum swelling degree between dry and swollen state, as it is also reproduced in the example \newline 
	\texttt{Q:"If active hydrogels should be applied for the opening and closing of channels, an adequate performance indicator is " \newline
	A:"the swelling ratio. The swelling ratio is defined as the ratio of the swollen gel volume to the dry..."\footnote{Generated with Llama-3.2-3B, Total output tokens: 20, argmax selection.}}
\end{enumerate}
While this list is focused on the knowledge for the separate steps, it is also relevant how the connection between the steps are represented in the model. For all these steps, specialized models exist in literature, which can be leveraged to check the accuracy of the LLM knowledge. For example, for the Processing $\to$ Structure step: To generate adequate output for this connection, \textit{context} for processing steps and the involved raw materials must be provided. In the Structure$\to$Properties step, the terminology of structure description can be very specialized and concise (which leads to its own problems, as will be discussed in section \ref{sec:res_unique_elements}).

\subsection{About generative pretrained transformer models}\label{sec:mat_met_llm_statistics}
The current work is focused on decoder-only Transformer models, since they are currently the most common architecture type of openly available large language models. Only the base models are considered: These are general-purpose foundation models that can be adapted for the use in many contexts, such as engineering and science, if they are published under an appropriate license. Hence, they are not instruction tuned (such as models used for chatbots) and are not trained to invoke tools like web-search or database search (such as agents/assistants or Retrieval-Augmented Generation systems). Instead, they generate new tokens based on the context. 

The process of generating text with these kinds of models is based on various steps. The prompt is divided into tokens\footnote{These tokens are usually used for billing of text generation through APIs of commercial LLMs, but they also play an important role in setting the context of the generation of new tokens.} according to a vocabulary. A word may be part of the vocabulary or subtokenized into parts. The tokens are mapped to vector embeddings that represent their meaning, as learned from the training data. In the next steps, positional encoding and the masked (causal) self-attention layers are applied to build a contextualized embedding of the sequence. The feed-forward layers of the model compute the final hidden states. From this, the logits are derived via the output projection, and the logits are then translated via the softmax function into a probability distribution for every possible token from the vocabulary. The next token is chosen by stochastic sampling according to the temperature parameter (a temperature of 1 leaves the probability distribution intact, temperature $<$ 1 sharpens it and temperature $>$ 1 flattens it), whereas a temperature of zero leads to the most probable token to be generated via the argmax function. 

If this process is viewed in context of unique material fingerprints, as discussed above, the following implications arise: A material name can be either part of the vocabulary, which makes their embedding more specialized than a series of sub-word tokens. However, if a material name is present in the vocabulary, this means that it has emerged as important enough in the training dataset to warrant its own token, according to the training algorithm. This can also mean that the word is used out of scientific/ engineering context, as will be discussed for chemical element names in section \ref{sec:res_unique_elements}. 

Please note that other architectures can be better suited for the tasks that are discussed here: For example, encoder-models like BERT \cite{Devlin2019} provide more suitable results for the embeddings and thus semantic similarity, which could be leveraged for judging similarity between materials according to their fingerprints. This can be evaluated by comparing significant dimensions in the embedding space. Furthermore, dimension reduction methods can be applied to extract new material-specific dimensions, since the subset of embeddings with material names is special in their context. 

However, the goal of the current work is to discuss the applicability of exactly these common models in engineering and science context. Furthermore, the availability is important, which lead to the choice of models from the largest providers that is given at the beginning of section \ref{sec:results_and_discussion}. 

If models generate tokens that are highly probable in embedding space, but have no counterpart in the real world, these are hallucinations \cite{Huang2025}. A low rate of hallucinations is important for the trustworthiness of a model in any aspect of application \cite{Wang2023}. 

\subsection{About chemical data and the Periodic Table of Elements (PTE)}\label{sec:mat_met_pse}
Chemical elements, as the building blocks of all materials, are the most basic ingredient of material knowledge in the discovery process. Every numerical result that is generated by a LLM in a material discovery reasoning process will include the context of chemical elements and their unique properties. 

It therefore makes sense to verify the truthfulness of data for the Periodic Table of Elements (PTE). Beside the classical arrangement by Mendeleev \cite{Mendelejew1869}, there are various other forms of arrangements for the chemical elements \cite{Schwerdtfeger2020}, which highlight for example the scarcity \cite{Sheehan1976}, etc. One output of the current work will be a PTE of material knowledge. In the current work, the melting temperature of pure elements (elementary substances), i.e., materials made of only one chemical element, is considered. The \texttt{mendeleev} Python package \cite{Mentel2021} provides an API to access the numerical data of the chemical elements for the truthfulness research in section \ref{sec:res_property_prediction}.

\clearpage
\section{Results and Discussion} \label{sec:results_and_discussion}
In the current section, different aspects of materials knowledge in LLMs are discussed. In section \ref{sec:res_unique_elements}, the uniqueness of tokens and their relevance is considered. This is followed by the presentation of the truthfulness in material data generation for LLMs of different sizes. 

\paragraph{Overview of the applied models}
The following open source large language models were applied in the current work, as shown in Table \ref{tab:res_model_tokenizer_info}. Please note that the model landscape is rapidly changing; the current status of open source models can be checked, e.g., at various leaderboards on Huggingface.co. The criteria for choosing models were as follows:
\begin{itemize}
	\item Availability: The models are available under permissive licenses, such as the MIT license (Microsoft phi-4), the Llama 3.1 Community License (Meta Llama), Apache-2.0 (Qwen, Mistral), or the gemma license (Google gemma). 
	\item Control: The models are required to be under full control by the scientist. There should be no wrappers or additional post-processing/filtering steps that change the output. Furthermore, the temperature is set to $0$, as substantiated in section \ref{sec:mat_met_llm_statistics}. These requirements are relevant for repeatability, however they limit the applicability of this study in everyday context: If scientists use close-source LLMs from providers such as OpenAI, Inflection, ..., then these parts are not under control. 
	\item Size: The current research was performed on a low-power computing system, which is realistic for many scientists/engineers that aspire to apply LLMs in their research, without resolving to supercomputer capabilities. The presented approach can be extended to include LLMs of any size, as long as the control and availability is ensured. 
\end{itemize}

\begin{table}[htbp]
	\centering
	\caption{Specifications of the Large Language Models that were investigated in the current work.}
	\label{tab:res_model_tokenizer_info}
	\begin{tabular}{@{}llr@{}}
		\hline
		Model & Parameters & Vocab Size \\
		\hline
		gemma-3-4b-pt \cite{Kamath2025} & 4B & 256000 \\
		gemma-3-1b-pt \cite{Kamath2025} & 1B & 256000  \\
		gemma-2-2b \cite{Riviere2024} & 2B & 256128 \\
		phi-4 \cite{Abdin2024} & 14B & 100352  \\
		Llama-3.1-8B \cite{Dubey2024} & 8B & 128000  \\
		Llama-3.2-3B \cite{Dubey2024} & 3B & 128000 \\
		Qwen3-4B \cite{Yang2025a} & 4B & 151669  \\
		Qwen3-1.7B \cite{Yang2025a} & 1.7B & 151669  \\
		Qwen3-0.6B \cite{Yang2025a} & 0.6B & 151669  \\
		Mistral-7B-v0.3 \cite{Jiang2023} & 7B & 32000 \\
		\hline
	\end{tabular}
\end{table}

The chosen models are only a small snapshot of all the materials that exist. The authors provide the code for benchmarking any other model of this architecture based on the same dataset and rules.

\subsection{Uniqueness of chemical element names} \label{sec:res_unique_elements}
As discussed in section \ref{sec:mat_met_llm_statistics}, tokenizing is not only relevant for billing, but also for representing information in the semantic space. Unique tokens for chemical elements ensure the immediate generation in a sequence. The alternative, i.e., subword tokenziation, instead leads to attention-based composition, which can be interrupted due to context and randomness (temperature). 

\begin{figure}[h]
	\centering
	\begin{subfigure}[b]{0.5\linewidth}
		\includegraphics[width=\linewidth]{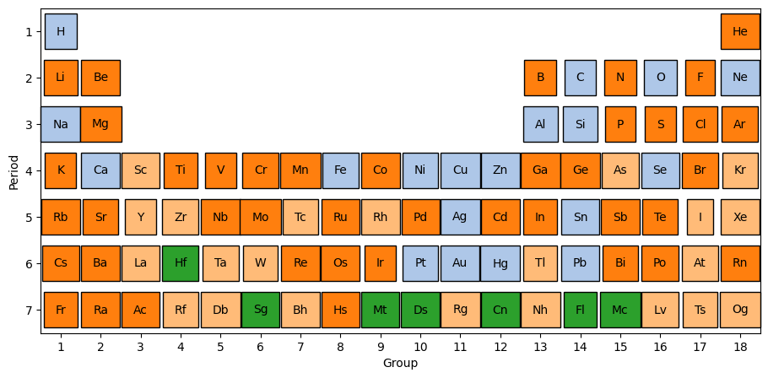}
		\caption{Number of tokens for chemical element names in gemma-3-1b-pt. }
	\end{subfigure} \quad
	\begin{subfigure}[b]{0.4\linewidth}
		\includegraphics[width=0.95\linewidth]{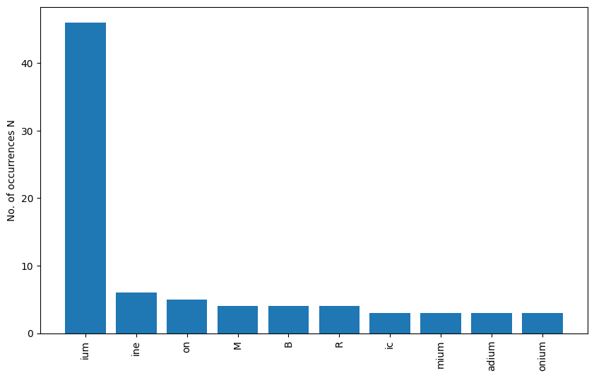}
		\caption{Most common subword tokens for elements in gemma-3-1b-pt.}
	\end{subfigure} \\
	\begin{subfigure}[b]{0.5\linewidth}
		\includegraphics[width=\linewidth]{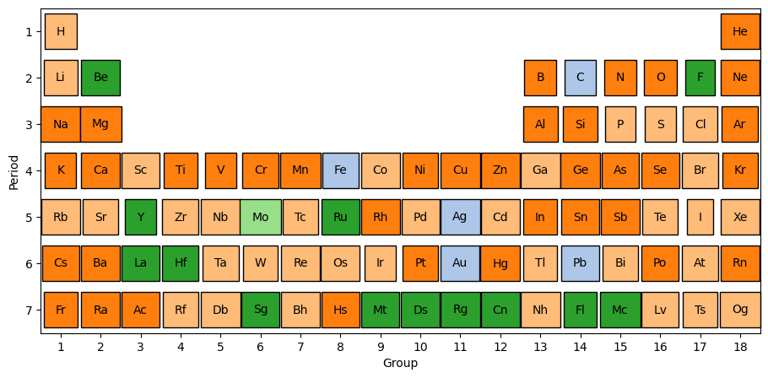}
		\caption{Number of tokens for chemical element names in Llama-3.2-3B.}
	\end{subfigure} \quad
	\begin{subfigure}[b]{0.4\linewidth}
		\includegraphics[width=0.95\linewidth]{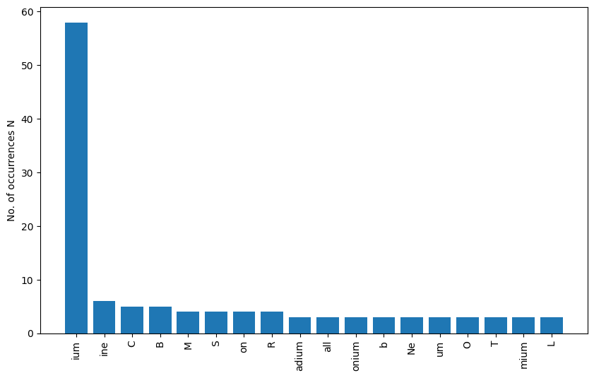}
		\caption{Most common subword tokens for elements in Llama-3.2-3B.}
	\end{subfigure} \\
	\begin{subfigure}[b]{0.8\linewidth}
		\includegraphics[width=\linewidth]{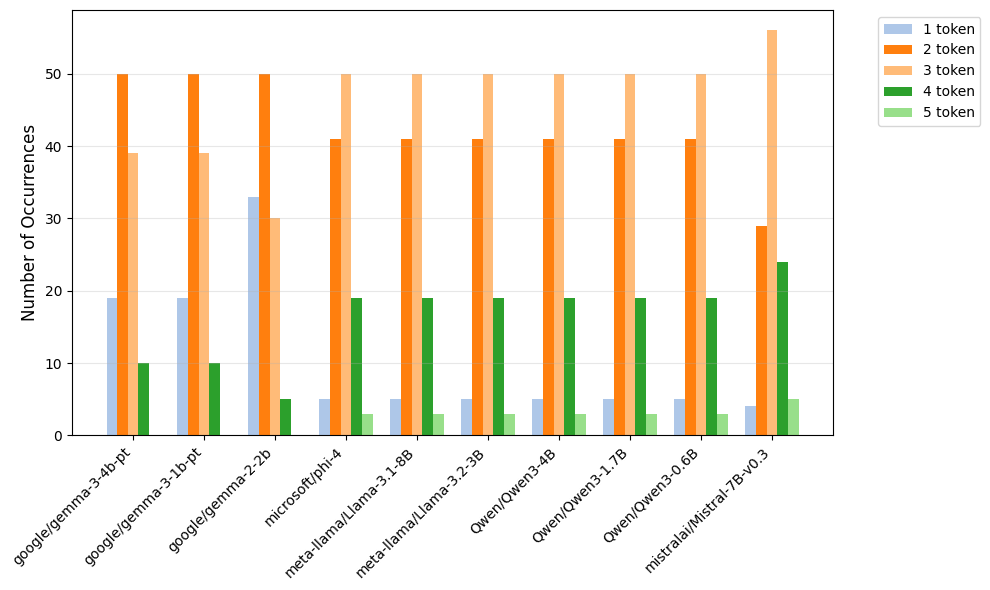}
		\caption{Number of tokens for each element name (including actinoid series and lanthanoid series). }
		\label{fig:res_uniqueness_list}
	\end{subfigure}
	\caption{Depiction of the periodic table of elements with the number of tokens for each element name. Results for the other models from Table \ref{tab:res_model_tokenizer_info} are shown in the appendix Figure \ref{fig:res_uniqueness_additional}.}
	\label{fig:res_uniqueness}
\end{figure}

In Figure \ref{fig:res_uniqueness}, results for the uniqueness of chemical element names in the vocabulary of every model from Table \ref{tab:res_model_tokenizer_info} are shown. Please note that the actinoid and lanthanoid series are not depicted here. To obtain these results, the element names were extracted from the mendeleev packages, tokenization was performed and the number of tokens (excluding special tokens) was counted. 

From the comparison of vocabulary terms in Figure \ref{fig:res_uniqueness_list}, differences between the models are visible. The gemma models show a higher count of unique tokens for chemical element names, which can be attributed to design of the vocabulary and of the training datasets \cite{Kamath2025}. 

The presented overviews show that common materials that have the same name as the chemical elements -- such as iron or silver -- always have a unique token, even in small vocabularies. It can be assumed that the additional uses of the names in the internet (in non-technical context) lead to a combination of meanings for the chemical element name: For example, "iron will" and "silver bullet" will increase the value in embedding dimensions that represent non-scientific concepts. 

This has implications for the unique material identifier (material name, fingerprint), which was discussed in section \ref{sec:mat_met}: Material names must be adequately specific such that they are not confused with trivial names. Furthermore, these identifiers must be added to the vocabulary of a specialized LLM (a fine-tuned version of a foundation model\footnote{General-purpose models can usually be fine-tuned in two ways to increase performance with respect to the model size: By choosing a language and by choosing a \textit{vertical}, such as science, legal, health, etc. Models for the application in PSPP chain reasoning should be specialized to English language because of its prevalence in science and then further adapted to material fingerprints.}), such that no subword tokenization is done on the material fingerprint. The dataset for fine-tuning must include a sufficient number of data points that substantiate the embeddings of these special tokens. This ensures that the respective embedding is specific in the materials dimension, which can lead to a better prediction of numerical values, as will be discussed in the following section.

\clearpage
\subsection{Accuracy of property value prediction in the PSE} \label{sec:res_property_prediction}
\begin{figure}[h]
	\centering
	\begin{subfigure}[b]{0.5\linewidth}
		\includegraphics[width=\linewidth]{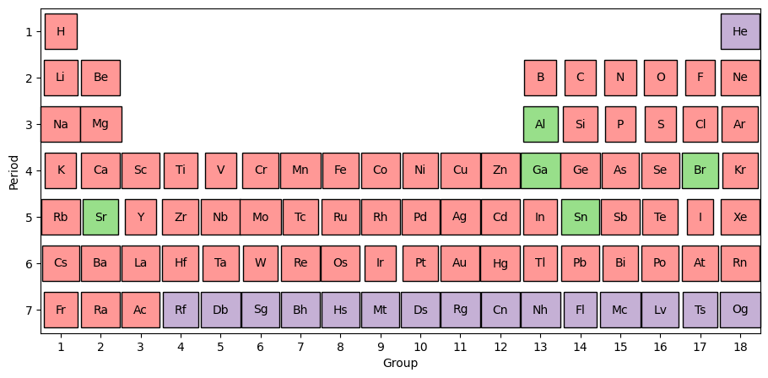}
		\caption{Truthfulness of generated data from gemma-3-1b-pt.}
	\end{subfigure} \quad
	\begin{subfigure}[b]{0.4\linewidth}
		\includegraphics[width=0.95\linewidth]{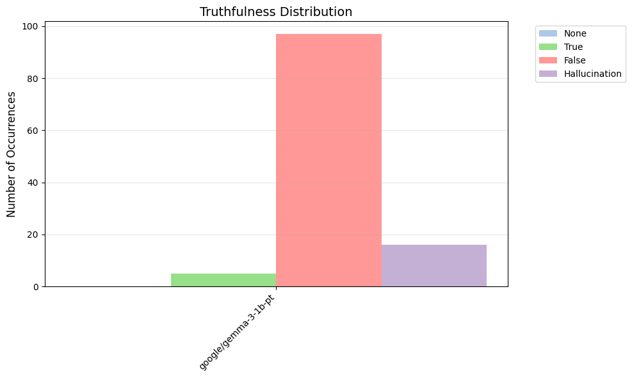}
		\caption{Overview for gemma-3-1b-pt.}
	\end{subfigure} \\
	\begin{subfigure}[b]{0.5\linewidth}
		\includegraphics[width=\linewidth]{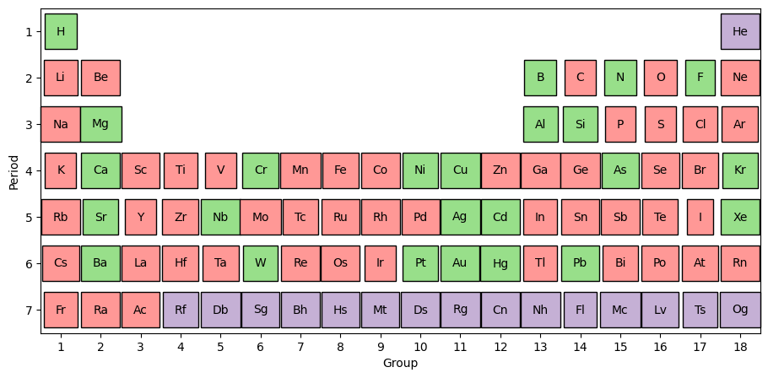}
		\caption{Truthfulness of generated data from gemma-2-2b.}
	\end{subfigure} \quad
	\begin{subfigure}[b]{0.4\linewidth}
		\includegraphics[width=0.95\linewidth]{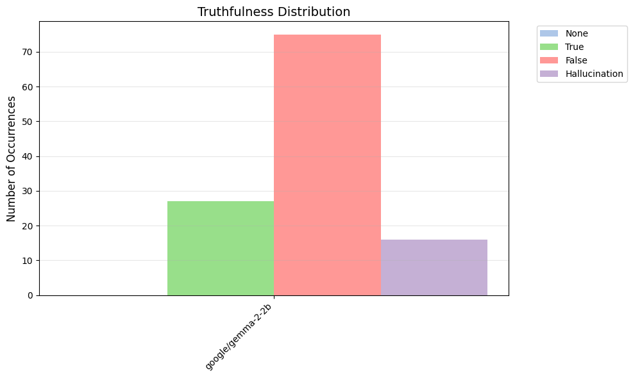}
		\caption{Overview for gemma-2-2b.}
	\end{subfigure} \\
	\begin{subfigure}[b]{0.5\linewidth}
		\includegraphics[width=\linewidth]{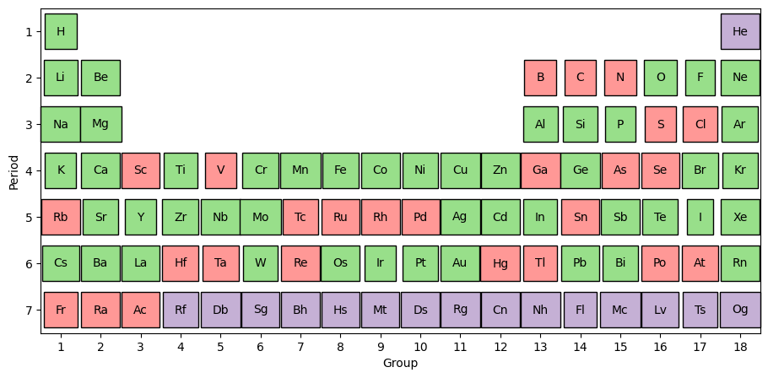}
		\caption{Truthfulness of generated data from llama-3.2-3B.}
	\end{subfigure} \quad
	\begin{subfigure}[b]{0.4\linewidth}
		\includegraphics[width=0.95\linewidth]{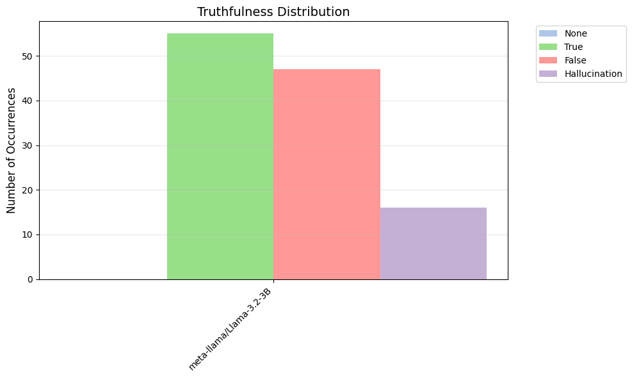}
		\caption{Overview for llama-3.2-3B.}
	\end{subfigure} \\
	\caption{Depiction of the periodic table of elements with the truthful prediction of the melting point.}
	\label{fig:res_truthfulness}
\end{figure}

In Figure \ref{fig:res_truthfulness}, the prediction results for the melting temperature of elementary substances (see section \ref{sec:mat_met_pse}) for different models is depicted. The process included the generation of an adequate prompt, the generation of output sequences, and the detection of correct values (also considering different unit systems through the use of regular expressions). Preliminary studies have shown that the prompt \texttt{"Pi is 3.14<. The melting temperature of ice in Kelvin is 273.15K<."} reliably allows the generation of easy-to-process model outputs: Through the concept of in-context learning, the number format, the unit and a custom end-of-sequence marker \texttt{<.} are established. 

From the results of the three models with 1, 2, and 3 billion parameters, it is visible that bigger models also enable a better reproduction of values. However, prediction of nonsensical values does not change: A melting temperatures of Helium is always predicted, even though at normal pressure, it remains gas or liquid. Furthermore, a melting point is also predicted for materials that are too short-lived for obtaining this kind of experimental result. These values are hallucinations and adequate prompting techniques (such as adding \texttt{"If no value exists, provide NaN."} or in-context learning \texttt{"The melting temperature of Helium is NaN."}) can lead to a reduction of falsely predicted values. Applying similar checks for all properties of the \texttt{mendeleev} package on all open source models enables the quantification of correct material knowledge about chemical elements and thus the informed choice of a foundation model for fine-tuning, which will be presented in future works. Furthermore, this can also lead to the outcome that certain general-purpose models are already capable enough to be included in the PSPP chain reasoning process.

\section{Conclusion} \label{sec:conclusion}
The current work is built around the question, how Large Language Models (LLMs) can be included in the material development and engineering design process. To answer the questions if these models can generate correct information about materials, the reproduction of data from the Periodic Table of Elements, was investigated. 

A discussion of the uniqueness of tokens leads to considerations about the design of material fingerprints with appropriate embeddings. These must be considered when creating specialized (fine-tuned) models for scientific/engineering application.

Furthermore, the generation of properties of elemental materials has shown that already for small models with 3 billion parameters, correct data can be present. However, the problem of hallucinations is easily visible and adequate mechanisms for counteracting this must be included in an end-to-end engineering model. 

In the current research environment for machine learning, it is hard to balance being critical about the application of LLMs for replacing well-founded physical models and the concern of missing opportunities for the own field. The current work can serve as a guide to material scientists and engineers on how to approach the inclusion of LLMs in their research software. To ensure truthful outcomes of an automated design process -- which can be realized as a Processing $\to$ Structure $\to$ Property $\to$ Performance chain reasoning process -- robust ways to quantify the correctness of LLMs over all possible materials are required. The current work is a starting point of building a benchmark for end-to-end tasks in materials science and engineering. By providing a mode of comparison for verifying results for every step in the PSPP chain, new and exciting models can be evaluated in their application scope and usefulness. 
e author.

\bibliographystyle{MSP}
\bibliography{../../my_mindmaps_and_references/Bibliogprahie_Habilitation}

\clearpage

\appendix
\section{Additional model results}

\begin{figure}[h!]
	\centering
	\begin{subfigure}[t]{0.5\linewidth}
		\includegraphics[width=\linewidth]{uniqueness_gemma_3_1b.png}
		\caption{Number of tokens for chemical element names in gemma\_3\_1b}
	\end{subfigure}
	\begin{subfigure}[t]{0.4\linewidth}
		\includegraphics[width=0.95\linewidth]{uniqueness_gemma_3_1b_tokens.png}
		\caption{Most common subword tokens for elements in gemma\_3\_1b}
	\end{subfigure} \\
	\begin{subfigure}[t]{0.5\linewidth}
		\includegraphics[width=\linewidth]{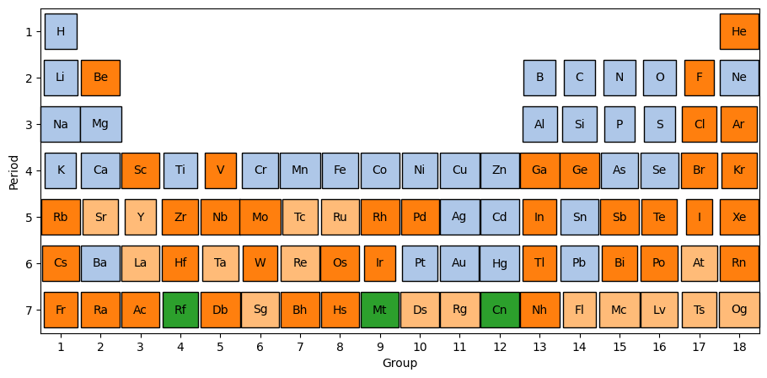}
		\caption{Number of tokens for chemical element names in gemma\_2\_2b}
	\end{subfigure} \quad
	\begin{subfigure}[t]{0.4\linewidth}
		\includegraphics[width=0.95\linewidth]{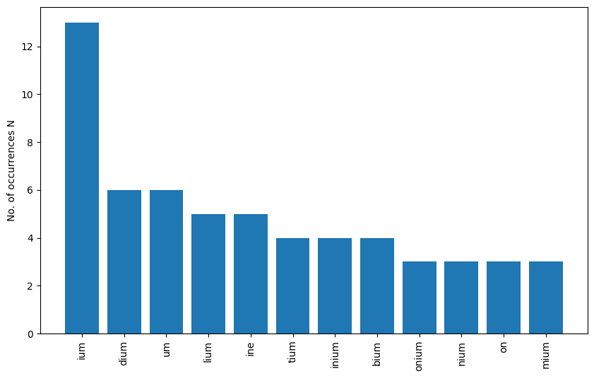}
		\caption{Most common subword tokens for elements in gemma\_2\_2b}
	\end{subfigure} \\
	\begin{subfigure}[t]{0.5\linewidth}
		\includegraphics[width=\linewidth]{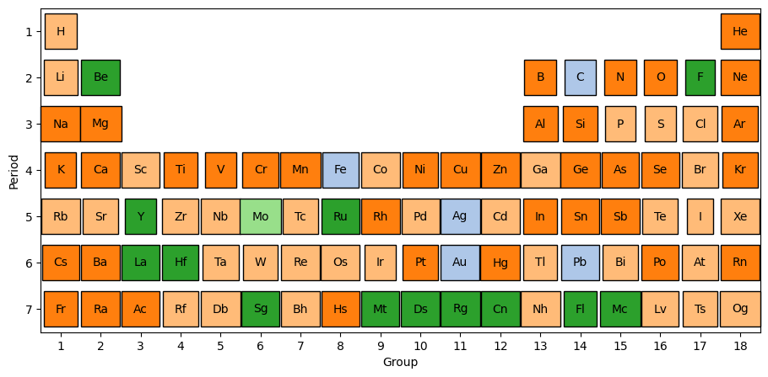}
		\caption{Number of tokens for chemical element names in phi-4}
	\end{subfigure} \quad
	\begin{subfigure}[t]{0.4\linewidth}
		\includegraphics[width=0.95\linewidth]{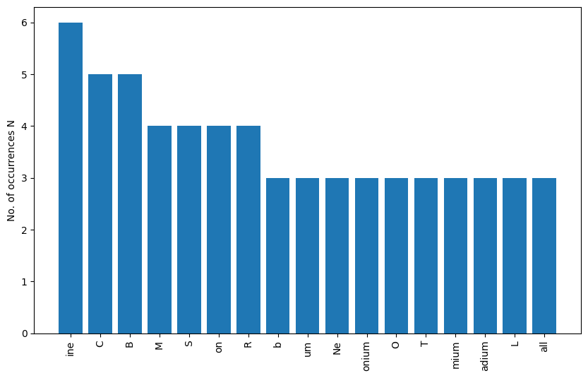}
		\caption{Most common subword tokens for elements in phi-4}
	\end{subfigure} \\
	\begin{subfigure}[t]{0.5\linewidth}
		\includegraphics[width=\linewidth]{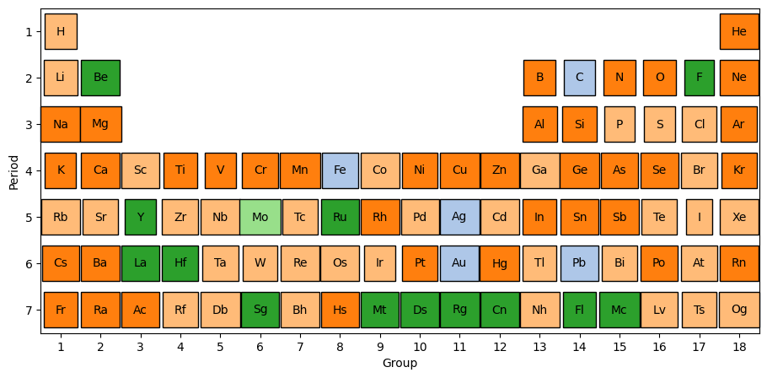}
		\caption{Number of tokens for chemical element names in Qwen3-4B}
	\end{subfigure} \quad
	\begin{subfigure}[t]{0.4\linewidth}
		\includegraphics[width=0.95\linewidth]{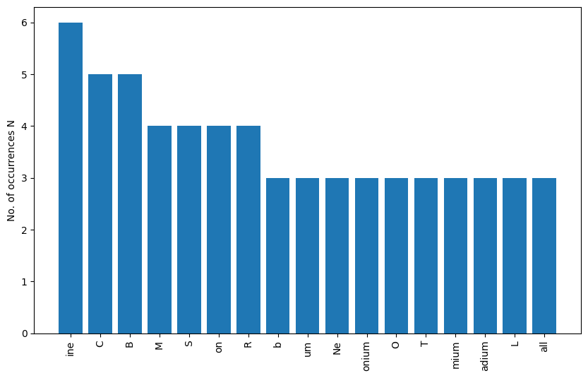}
		\caption{Most common subword tokens for elements in Qwen3-4B}
	\end{subfigure}
	\caption{Additional results for uniqueness of tokens (see Figure \ref{fig:res_uniqueness}).}
	\label{fig:res_uniqueness_additional}
\end{figure}

\end{document}